\documentclass[12pt,preprint]{aastex}

\usepackage{subfigure}

\shorttitle{Characteristics of Anemone Active Regions}

\begin{document}

\title{Characteristics of Anemone Active Regions Appearing in Coronal
Holes Observed with {\it Yohkoh} Soft X-ray Telescope}
\shortauthors{Asai et al.}

\author{
Ayumi Asai\altaffilmark{1,2,3},
Kazunari Shibata\altaffilmark{4},
Hirohisa Hara\altaffilmark{2,3}, and
Nariaki V. Nitta\altaffilmark{5}}

\email{asai@nro.nao.ac.jp}

\altaffiltext{1}{
Nobeyama Solar Radio Observatory, National Astronomical Observatory of
Japan, Minamimaki, Minamisaku, Nagano, 384-1305, JAPAN}

\altaffiltext{2}{
National Astronomical Observatory of Japan, Osawa, Mitaka, Tokyo,
181-8588, JAPAN}

\altaffiltext{3}{
The Graduate University for Advanced Studies (SOKENDAI), JAPAN}

\altaffiltext{4}{
Kwasan and Hida Observatories, Kyoto University, Yamashina, Kyoto,
607-8471, JAPAN}

\altaffiltext{5}{
Lockheed Martin Solar and Astrophysics Laboratory, Department/ADBS,
B/252, 3251 Hanover Street, Palo Alto, CA 94304, U.S.A.}


\begin{abstract}
Coronal structure of active regions appearing in coronal holes is
studied by using the data obtained with the Soft X-Ray Telescope (SXT)
aboard {\it Yohkoh} from 1991 November to 1993 March.
The following characteristics are found; 
Many of active regions appearing in coronal holes show a structure that
looks like a ``sea-anemone''.
Such active regions are called {\it anemone ARs}.
About one-forth of all active regions that were observed with SXT from
their births showed the anemone structure.
For almost all the anemone ARs, the order of magnetic polarities is
 consistent with the Hale-Nicholson's polarity law.
These anemone ARs also showed more or less east-west asymmetry in X-ray
intensity distribution, such that the following (eastern) part of the
ARs is brighter than its preceding (western) part.
This, as well as the anemone shape itself, is consistent with the
magnetic polarity distribution around the anemone ARs.
These observations also suggest that an active region appearing in
coronal holes has simpler (less sheared) and more
preceding-spot-dominant magnetic structure than those appearing in
other regions.
\end{abstract}



\section{Introduction}
The Soft X-ray Telescope (SXT: Tsuneta et al. 1991) aboard {\it Yohkoh}
\citep{Oga91} enabled us to study detailed structure and evolution of
coronal part of active regions (ARs).
Among many findings, one of the interesting discoveries is
``sea-anemone'' like structure (Shibata et al. 1994a, 1994b; see also
Fig. 1).
This phenomenon is characterized by radially aligned coronal loops that
connect the opposite magnetic polarity of the AR magnetic field and the
surrounding region with the unipolar field, and is physically the same as
X-ray ``fountain'' originally reported by \citet{Tou73} and
\citet{Shee75a} in the {\it Skylab} era.
We call these active regions the {\it anemone AR} in this paper.
The appearance of anemone ARs typically lasts for a few days.
Figure 2 and 3 present examples of anemone AR evolutions.
Some ARs stably show anemone structure for a couple of weeks as shown in
Figure 2, and others show anemone structure for a period of the
evolution like as Figure 3.
Figure 2 also shows a jet activity.
Extreme ultraviolet Imaging Telescope (EIT; \cite{Dela96}) on board the
Solar and Heliospheric Observatory (SOHO; \cite{Dom95}) have shown
similar features in the extreme ultraviolet images.
The appearance is almost the same as those seen in SXT.

Anemone ARs are thought to often appear in coronal holes (CHs) that
consist of global open magnetic field.
A bipole emerging within a CH magnetically reconnects with the CH field
to produce the characteristic structure (Sheeley et al. 1975b; see also
Fig. 4).
This situation of the magnetic reconnection between the emerging flux
and the surrounding field is suitable to generate X-ray jets and/or
H$\alpha$ surges \citep{Yoko95,Yoko96}, and indeed many jets ejected
from anemone ARs have been observed \citep{Shi94b,Vour96,Kun99,Alex99}.
\citet{Wang06} and \citet{Nitta06} investigated the solar origins of
$^3$He-rich solar energetic particle (SEP) events, and found that the
sources of the impulsive SEPs lie next to CHs containing Earth-directed
open field lines.
These sources further showed the association with jetlike ejections seen
in extreme-ultraviolet images, and type III radio bursts.
\citet{Wang98} indicates the possibility that even polar plumes are
associated with jets from anemone ARs.
A small anemone AR in a CH probably evolves into a polar plume as one of
their flux that has the opposite magnetic polarity to the CH gradually
cancels out, and generates jets.
Anemone ARs sometimes generate filament eruptions \citep{Cher02}, and
even large flares and/or coronal mass ejections
\citep{Ver98,Liu06,Liu07,Asa07}.
For these cases, eruptions occurring in CHs easily travel keeping the
speeds by having an advantage of high speed solar wind from CHs.
Anemone ARs are probably related with non-radial coronal streamers
emanating from magnetically high latitudes \citep{Sai00}, and attentions
to their relation with fast solar winds have been paid
\citep{Taka94,Sai94,Wang98}.
Therefore, the understandings physical and morphological characteristics
of anemone ARs are important for space weather studies.

However, it has not been well understood how frequently such an
asymmetric magnetic configuration in bipolar spots occurs in CHs.
Meanwhile, the reason why the anemone ARs look like a sea-anemone has
been thought that the average magnetic field strength in one of bipolar
spots is stronger than that of another spot, and that magnetic polarity
of the stronger spot is opposite to the ambient polarity of CHs
(unipolar regions), while it should be confirmed by investigating the
characteristics of anemone ARs and CHs.
In this paper we statistically examine the features of anemone ARs
observed with {\it Yohkoh}/SXT.
We investigate the birth places of anemone ARs, and confirm their
relations with CHs.
Then, we investigate the characteristics of anemone ARs, such as the
magnetic configurations of emerging flux regions, that of the
surrounding CHs, how anemone ARs appear in CHs, and so on.
We also study the relation between anemone ARs and X-ray jets, since it
is suggested qualitatively, but has been unclear quantitatively.
In \S 2 we describe the observations and the results, and in \S 3 we
summarize our results and offer discussions.

\section{Observations and Results}
We used soft X-ray (SXR) images taken with SXT full frame images (FFI)
of SXT.
The spatial resolution of FFI images is either half resolution ($\sim$ 5
arcsec) or quarter resolution ($\sim$ 10 arcsec).
The time resolution  ranges from a few minutes to an hour.
The SXT filter used for the FFI events in this paper is either 0.1
$\mu$m Al filter or Al/Mg/Mn filter, both of which are sensitive to the
SXRs between 3 and 60 {\AA}.

There are 49 ARs whose births were observed with SXT from 1991 November
to 1992 May.
We examined the birth places of those ARs, and categorized them into the
following three types according to the appearances; A: anemone type, B:
two-sided-loops type, and C: other types.
We summarize the results in Table 1.
The two-sided-loops structure is large-scale loop brightenings that occur
at both sides of emerging fluxes.
The structure suggests a magnetic reconnection between the emerging flux
and an overlying coronal magnetic field that lies nearly horizontally
\citep{Shi94a}.
Figure 4 shows schematic illustrations of an anemone ARs and a
two-sided-loops structure.
A newly emerged magnetic flux interacts with the surrounding fields and
generates the characteristic configuration.
If the surrounding fields stand vertically, such as for a CH (the top
and middle panels of Fig. 4), the interaction leads to an anemone
structure.
On the other hand, a two-sided-loops configuration is generated, if the
surrounding fields lie nearly horizontally.
Among the newly emerged ARs, 12 ARs showed clear anemone structure.
Therefore, an anemone AR were not rare phenomena and about one-forth of
all newly emerged ARs belonged to the anemone ARs.
Ten of all anemone ARs (type A) appeared in CHs, and only one anemone AR 
appeared in QRs.
On the other hand, the ARs that appeared in QRs usually did not show
anemone structure, but mainly two-sided-loops.
The difference between magnetic field configurations of CHs with
nearly vertical fields and those of QRs with nearly horizontal fields
leads to the difference of the appearances.
Table 1 also presents the association of anemone ARs with X-ray jets.
About 58~\% of Anemone ARs (7/12) showed jetlike ejections, and
therefore, we can confirm that anemone ARs are suitable for X-ray jets
as \citet{Shi94b} reported.

We also sought anemone ARs in the SXT/FFI images between November 1991
and March 1993, and found 28 anemone ARs.
All of these anemone ARs appeared within CHs.
The anemone AR is defined by the following rules;
(1) It shows a configuration of loops fanning almost symmetrically in
the SXT images.
(2) It looks an isolated active region in the SXT images.
A typical example of anemone AR, which was observed on 1992 January 10,
is presented in Figure 1.
Figure 1 shows the a SXR image (panel a) and a visible light image
(panel b) taken with {\it Yohkoh}/SXT, and a magnetogram (panel c)
obtained with Kitt Peak, National Solar Observatory.
Figure 1a clearly shows that many loops fan out symmetrically from the
center of the anemone.
We can also see following-preceding (east-west) asymmetry in the
brightness of loops; in this case, loops in the following (eastern) part
of the AR is brighter than those of the preceding (western) part.
Table 2 lists all the 28 anemone ARs with their SXR and magnetic field
characteristics.
The column 4 of Table 2 show the following-preceding asymmetry.
We can recognize that almost all of the anemone ARs show the asymmetry.
By using full disk magnetograms taken at Kitt Peak, National Solar
Observatory, we further statistically studied the characteristics of the
anemone ARs, such as the magnetic polarities of surrounding CHs (column
5), and those of bipolar spots/regions (column 6).
We also added the sunspot magnetic classification from Solar Geophysical
Data (column 7) to Table 2.

From Table 2, we found some characteristics of anemone AR.
First of all, although anemone ARs mainly showed simple structure, they
were not always $\alpha$-type sunspots.
$\beta$-type or even more complex sunspots can also generate anemone
structure, by interacting with the surrounding magnetic field after the
emergence.
Then, we found that 71~\% of all anemone ARs (20/28) appeared in the
northern hemisphere.
The sunspot number in this period was larger in the southern hemisphere,
and therefore, anemone ARs showed a tendency of the anti-solar activity.
This is consistent with the fact that almost all anemone ARs appeared
within CHs.
Almost all the surrounding CHs in the northern hemisphere had the
positive magnetic polarity during this period.
Among the anemone ARs in the northern hemisphere, about 80~\% (16/20)
had the characteristics that magnetic polarity of the preceding spots
was negative, which was, therefore, opposite to that of the the
surrounding CHs.
We call these ARs {\it normal} anemone ARs (see the left panel of
Table~3).
The order of the magnetic polarities for most of the normal anemone,
that is, negative (positive) polarity for the preceding (following)
spots, is consistent with the Hale-Nicholson's polarity law
\citep{Hal19} in the northern hemisphere.
On the other hand, only four anemone ARs (20~\%) had different magnetic
configurations; there was one anemone in which both the the polarity of
the preceding and that of the CH were positive, and only three anemone
ARs that appeared in CHs with the negative magnetic polarity (see
Table~2).
We call these cases {\it abnormal} anemone ARs.
These anemone ARs also showed more or less east-west asymmetry in SXR
intensity distribution.
We examined the relation between the normal/abnormal features with the
following-preceding asymmetry.
75~\% of all normal anemone ARs (15/20) showed a clear tendency that the
following (eastern) part of the ARs was brighter than their preceding
(western) part.
We summarize these features in Table~3, according to the magnetic
polarities (the left panel) and to the asymmetry of the loop brightness
(the right panel).

In the southern hemisphere, on the other hand, the magnetic polarity of
preceding spots of all anemone ARs were positive, although the total
number of anemone ARs was much smaller (8).
The order of the magnetic polarities for the preceding/following spots
is again consistent with Hale-Nicholson's polarity law in the southern
hemisphere.
Interestingly, most of them (7/8) appeared in CHs with the positive
magnetic polarity.
During the period from 1991 November to 1993 March, there were both CHs
with positive and those with negative polarities on the solar surface.
Therefore, we can say that anemone ARs tend to occur in CHs with the 
positive magnetic polarity in this period.
The preceding (western) part of loops were dominantly brighter in SXRs
than the following part in half (4/8) of these ARs.
We also summarize these in Table~3.

\section{Summary and Discussions}
We statistically studied the characteristics of anemone ARs observed
with {\it Yohkoh}/SXT.
First, we surveyed 49 ARs whose births were observed with SXT from 1991
November to 1992 May, and found the following feature;
(1) About one-forth of all newly emerged ARs (12/49) were anemone ARs.
Moreover, almost all anemone ARs appeared within CHs, and the ARs that
appeared in QRs did not show anemone structure but mainly two-sided-loops.
We also confirmed that anemone ARs usually generate X-ray jets.
Next, we examined 28 anemone ARs observed between 1991 November and 1993
March.
The following characteristics were found; 
(2) About 71~\% of all anemone ARs appeared in the northern hemisphere,
while the sunspot number was larger in the southern hemisphere, in this
period.
This means that the number of anemone ARs has a feature of the
anti-solar activity.
Furthermore, almost all anemone ARs were not located on or near the
global neutral line where active longitudes were situated,
(3) Among the anemone ARs in the northern hemisphere, about 80~\% had
the characteristics that magnetic polarity of the preceding spots was
negative, which is consistent with the Hale-Nicholson's polarity law.
It is opposite to that of surrounding CHs, since almost all the
surrounding CHs in the northern hemisphere had the positive magnetic
polarity during this period.
In the southern hemisphere, on the other hand, the preceding spots of all
anemones was positive, which is consistent with the Hale-Nicholson's
polarity law.
The magnetic polarity of surrounding CHs was, interestingly, mainly
positive, while there are both CHs with positive and those with negative
polarities in the southern hemisphere during the period,
(4) Anemone ARs showed more or less following-preceding (east-west)
asymmetry in SXR intensity distribution.
Especially, the normal anemones in the northern hemisphere showed that
the following (eastern) part of the ARs was brighter than their
preceding (western) part.
For a half of anemone ARs (4 of all 8 ARs) in the southern hemisphere,
on the other hand, the preceding (western) part of loops were brighter
in SXRs than the following part.

The observational features (1), (2), and (3) suggest that the anemone
ARs have simpler (less sheared) magnetic structure than other ARs.
The ``anemone'' shape itself shows a potential like magnetic
configuration, that is, the lowest energy state.
This is also consistent with the facts that $\alpha$- $\beta$-type
spots, which are less active ARs, are observed at the center of anemone
ARs.
Moreover, the observed anemone ARs showed the preceding-spot-dominant
magnetic structure clearer than those appearing in other regions.
We also followed the evolution of anemone ARs, and found that a typical
anemone AR does not changes its appearance even when it approaches the
solar limb \citep{Sai00}.
The feature (4) showed that the following-preceding asymmetry in SXR
intensity distribution depends on the order of the magnetic polarities
of anemone ARs and CH.

We can see a clear tendency that anemone ARs appear within CHs with the
positive polarity.
However, we cannot conclude it, since the period studied in this paper
is restricted on a part of a solar-cycle (from 1991 to 1993).
In the northern hemisphere almost all CHs have the positive polarity,
and for CHs in the southern hemisphere, the number of observed anemone
ARs is too small during this period.
Moreover, although we found an anti-solar activity of anemone ARs, we
need more samples covering a longer-term before we conclude it.
We are required to survey anemone ARs with the data that cover one
solar-cycle observed with {\it Yohkoh}/SXT in future works.
The survey of anemone ARs covering one solar-cycle will also make clear
the variation of these features through the solar-cycle and the
association with fast solar winds.

We also have to make clear whether emerging fluxes that generate anemone
ARs as themselves have special characteristics or not, and how they
relate with the origin of magnetic fields.
More detailed examinations of anemone ARs and structure of the emerging
fluxes, by using data that have higher spatial resolution and
sensitivity are required to answer these questions.
For example X-Ray Telescope on board {\it Hinode} have observed many
similar (and smaller) features, and analyzing them will be appropriate.


\acknowledgments

We first acknowledge an anonymous referee for his/her useful comments 
and suggestions.
We wish to thank Prof. T. Saitoh for fruitful discussions and his
helpful comments.
This work was supported by the Grant-in-Aid for Creative Scientific
Research ``The Basic Study of Space Weather Prediction'' (17GS0208, Head
Investigator: K. Shibata) from the Ministry of Education, Science,
Sports, Technology, and Culture of Japan.
The {\it Yohkoh} satellite is a Japanese national project, launched and 
operated by ISAS, and involving many domestic institutions, with 
multilateral international collaboration with the US and the UK.



\begin{table}
\begin{center}
Table 1 \\
Birth place of active regions whose birth were observed by {\it
Yohkoh}/SXT from 1991 November to 1992 May\\
\begin{tabular}{lrrrrr}\tableline\tableline
Type & Total & QR & CH & QR/CH$^{a}$ & Jet$^{b}$ \\ \tableline
A (anemone)         & 12  &  1 & 10 &  1 &  7 \\
B (two-sided-loops) & 13  & 13 &  0 &  0 &  9 \\
C (others)          & 24  & 18 &  5 &  1 &  7 \\ \tableline
\end{tabular}
\end{center}
$^{a}$ Boundary between QR and CH.\\
$^{b}$ The number of ARs that were associated with jet or jet-like
phenomena.\\
\end{table}

\begin{table}
\begin{center}
Table 2 \\
List of anemone ARs with their SXR and magnetic characteristics\\
\begin{tabular}{lccccccl}\tableline\tableline
Date & NOAA AR & Helio. Lat.$^{a}$ & Carr. Rot.$^{b}$ & X-asym.$^{c}$ & 
CH pol. & AR pol.$^{c}$ & Mag. class$^{e}$  \\ \tableline
1991-Nov-10 & 6921 & N   & ~    & F & $+$ &  $+ -$  & B \\
1992-Jan-10 & 7001 & N25 & 1851 & F & $+$ &  $+ -$  & A \\
1992-Fev-07 & 7051 & N21 & 1852 & F & $+$ & ($+ -$) & B \\
1992-Mar-07 & 7085 & N23 & 1853 & F & $+$ &  $+ -$  & A \\
1992-Mar-07 & 7095 & N16 & 1853 & F & $+$ &  $+ -$  & A \\
1992-Apr-04 & 7124 & N14 & 1854 & F & $+$ &  $+ -$  & B \\
1992-Apr-25 & 7145 & N11 & 1855 & F & $+$ &  $+ -$  & B \\
1992-May-02 & 7146 & N08 & 1855 & F & $+$ &  $+ -$  & A \\
1992-May-21 & 7174 & N14 & 1856 & F & $+$ &  $+ -$  & B \\
1992-Jun-20 & 7205 & N11 & 1857 & F & $+$ &  $+ -$  & B \\
1992-Aug-15 & 7263 & N16 & 1859 & F & $+$ &  $+ -$  & B \\
1992-Sep-11 & 7276 & N15 & 1860 & F & $+$ &  $+ -$  & BGD \\
1992-Dec-19 & 7375 & N14 & 1963 & F & $+$ &  $+ -$  & B \\
1992-Dec-24 & 7381 & N07 & 1863 & F & $+$ &  $+ -$  & B \\
1993-Jan-26 & 7409 & N18 & 1865 & F & $+$ &  $+ -$  & B \\
1991-Dec-18 & 6973 & N   & ~    & unclear & $+$ &  $+ -$  & B \\
1992-Jan-31 & 7029 & N18 & 1852 & P & $-$ &  $+ -$  & B \\
1992-Jun-07 & 7192 & N09 & 1856 & P & $-$ &  $+ -$  & B \\
1991-Nov-12 & 6918 & N   & ~    & F & $-$ &  $- +$  & A \\
1993-Feb-09 & east of 7417 & N & ~ & unclear & $+$ & $- +$ & no data \\ \tableline
1991-Nov-20 & 6928 & S   & ~    & P & $+$ &  $- +$  & A \\
1992-Apr-25 & 7143 & S05 & 1854 & P & $+$ &  $- +$  & BGD \\
1992-May-02 & 7150 & S07 & 1855 & P & $+$ &  $- +$  & B \\
1992-May-16 & 7167 & S07 & 1855 & P & $+$ &  $- +$  & B \\
1992-May-21 & 7167new & S07 & 1856 & unclear & $+$ & $- +$ & B \\
1992-May-21 & 7176 & S12 & 1856 & unclear & $+$ &  $- +$  & A \\
1992-Jun-15 & 7195 & S08 & 1857 & F & $+$ &  $- +$  & A \\
1991-Nov-03 & 6900 & S   & ~    & F & $-$ &  $- +$  & A \\ \tableline
\end{tabular}
\end{center}
$^{a}$ Heliographic latitude from Solar-Geophysical Data
(http://www.ngdc.noaa.gov/stp).\\
$^{b}$ Carrington rotation number Solar-Geophysical Data.\\
$^{c}$ X-ray asymmetry.
F (P) means that the following (preceding) part of the anemone structure
is brighter.\\
$^{d}$ Magnetic polarities of anemone ARs.
The left and the right signs show the following and the preceding spots,
respectively.\\
$^{e}$ Sunspot magnetic classification from Solar-Geophysical Data.\\
\end{table}

\begin{table}
\begin{center}
Table 3 \\
Summary of anemone ARs according to the magnetic polarities ({\it left})
and to the asymmetry of the loop brightness ({\it right}) \\
\begin{tabular}{lr | lr}\tableline\tableline
magnetic pol.  &  & asymmetry & \\ \tableline
northern hemisphere    & 20 & northern hemisphere   & 20 \\
$\;\;$ CH $+$ AR $+ -$ (normal) & 16 & $\;\;$ following (eastern) bright X-loops & 16$^{a}$ \\
$\;\;$ CH $+$ AR $- +$ & 1 & $\;\;$ preceding (western) bright X-loops &  2 \\
$\;\;$ CH $-$ AR $+ -$ & 2 & $\;\;$ unclear                            &  2 \\
$\;\;$ CH $-$ AR $- +$ & 1 & & \\
southern hemisphere    & 8 & southern hemisphere   &  8 \\
$\;\;$ CH $+$ AR $+ -$ & 0 & $\;\;$ following (eastern) bright X-loops &  2 \\
$\;\;$ CH $+$ AR $- +$ & 7 & $\;\;$ preceding (western) bright X-loops &  4 \\
$\;\;$ CH $-$ AR $+ -$ & 0 & $\;\;$ unclear                            &  2 \\
$\;\;$ CH $-$ AR $- +$            &  1 & & \\ \tableline
total of anemone ARs   & 28 & total of anemone ARs  & 28 \\ \tableline
\end{tabular}
\end{center}
$^{a}$ 15 of them are normal type, and 1 is abnormal type.
\end{table}


\begin{figure}
\epsscale{.80}
\plotone{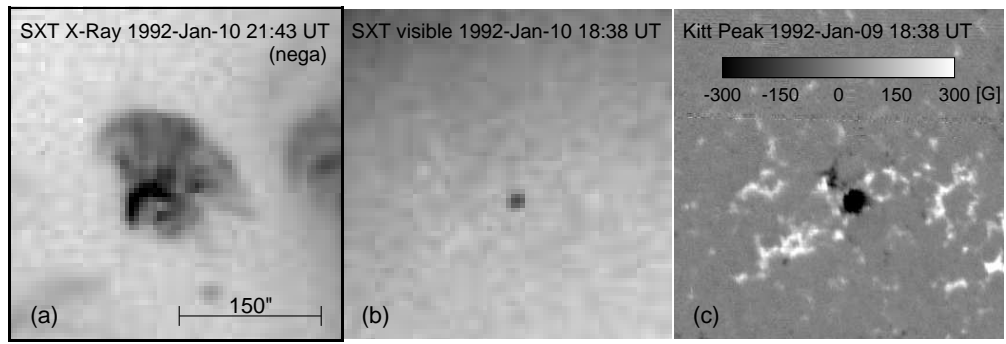}
\caption{
Typical example of the anemone AR observed on 1992 January 10 (NOAA AR
is 7001).
(a) SXR image taken with {\it Yohkoh}/SXT.
(b) Visible light image taken with the aspect sensor of SXT.
(c) Magnetogram taken with Kitt Peak.
}
\end{figure}

\begin{figure}
\epsscale{.40}
\plotone{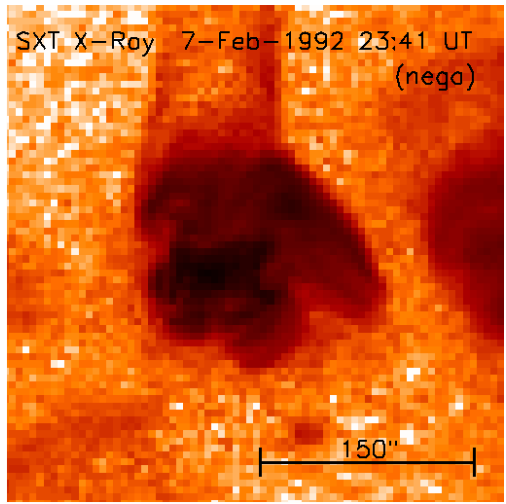}
\caption{
({\it for associated mpeg file :
 http://www.journals.uchicago.edu/action/showMedia?doi=10.1086\%2F523842\&type=video\_original\&id=video2.mpg})
Typical example of anemone AR evolution.
Movie of SXR images for NOAA AR 7001 shows a jet activity.
}
\end{figure}

\begin{figure}
\epsscale{.40}
\plotone{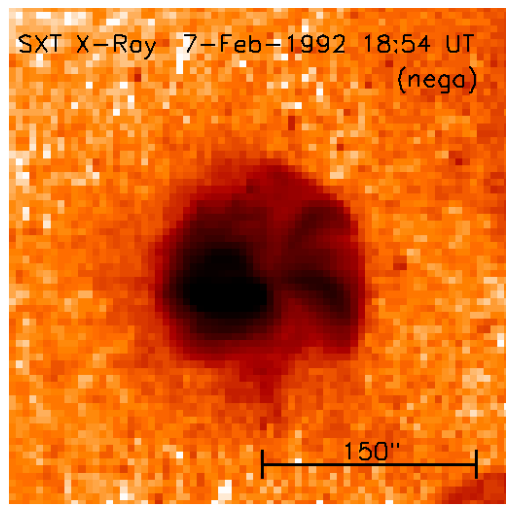}
\caption{
({\it for associated mpeg file :
 http://www.journals.uchicago.edu/action/showMedia?doi=10.1086\%2F523842\&type=video\_original\&id=video3.mpg})
Typical example of anemone AR evolution.
Movie of SXR images for NOAA AR 7051 shows that the anemone feature
appears in the course of the evolution.
}
\end{figure}

\begin{figure}
\epsscale{.80}
\plotone{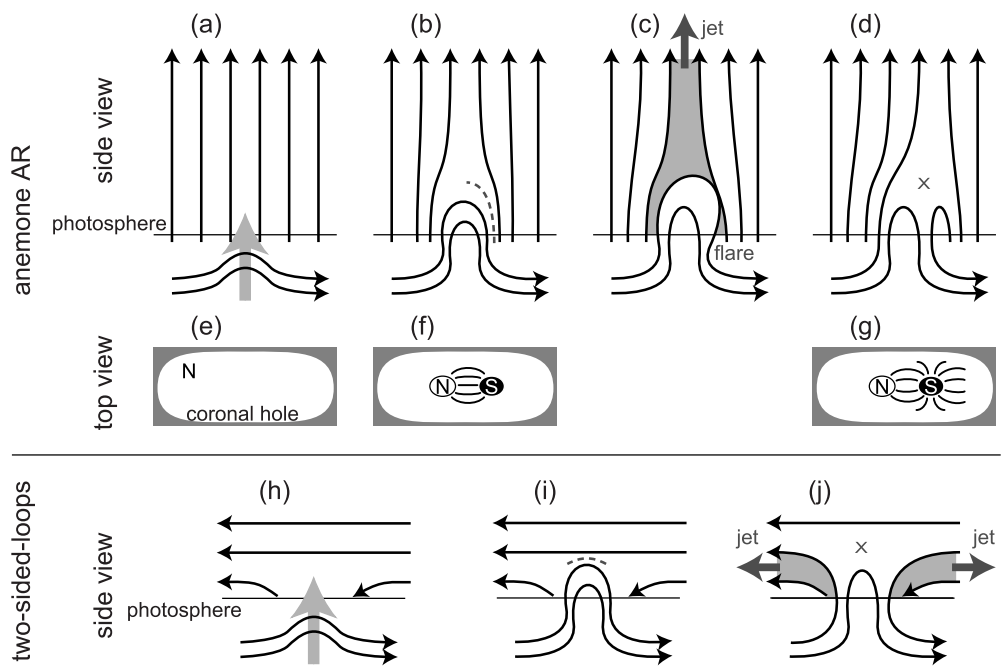}
\caption{
Schematic illustrations of an anemone AR in a side view (top panels) and
in a top view (middle panels), and those of a two-side-loops structure
in a side view (bottom panels).
(a): A magnetic flux newly emerges within a unipolar region such as a
CH.
(b): A current sheet is generated between the emerged field and the
surrounding field as shown with the dashed gray line.
(c): A magnetic reconnection occurs there, and generates jets and flare
brightenings.
(d) and (g): An anemone structure is formed.
The gray X-sign shows a X-point.
(h), (i), and (j): A magnetic flux newly emerges within a quiet region
that is consist of horizontal magnetic fields, and a magnetic
reconnection between them generates jets ejected both sides of X-point.
The shaded gray region in (c) and (j) is filled with hot plasma that
emits SXRs.
}
\end{figure}


\begin{thebibliography}{}
\bibitem[Alexander and Fletcher(1999)]{Alex99}
Alexander, D., and Fletcher, L. 1999, \solphys, 190, 167

\bibitem[Asai et al.(2007)]{Asa07}
Asai, A., Shibata, K., Ishii, T. T., Oka, M., Fujiki, K., Kataoka, R.,
Gopalswamy, N. 2007, submitted

\bibitem[Chertok et al.(2002)]{Cher02}
Chertok, I. M., Mogilevsky, E. I., Obridko, V. N., Shilova, N. S.,
Hudson, H. S. 2002, \apj, 567, 1225

\bibitem[Delaboudini\'{e}re et al.(1996)]{Dela96}
Delaboudini\'{e}re, J.-P. 1996, \solphys, 162, 291

\bibitem[Domingo, Fleck, and Poland(1995)]{Dom95}
Domingo, V., Flack, B., and Poland, A. I. 1995, \solphys, 162, 1

\bibitem[Hale et al.(1919)]{Hal19}
Hale, G. E., Ellerman, F., Nicholson, S. B., Joy, A. H. 
1919, \apj, 49, 153


\bibitem[Kundu et al.(1999)]{Kun99}
Kundu, M. R., Nindos, A., Raulin, J.-P., Shibasaki, K., White, S. M.,
Nitta, N., Shibata, K., Shimojo, M. 1999, \apj, 520, 391

\bibitem[Liu \& Hayashi(2006)]{Liu06}
Liu, Y. \& Hayashi, K. 2006, \apj, 640, 1135

\bibitem[Liu (2007)]{Liu07}
Liu, Y. 2007, \apj, 654, L171




\bibitem[Nitta et al.(2006)]{Nitta06}
Nitta, V. N., Reames, D. V., DeRosa, M. L., Liu, Y., Yashiro, S.,
Gopalswamy, Natchimuthuk 2006, \apj, 650, 438

\bibitem[Ogawara et al.(1991)]{Oga91}
Ogawara, Y., Takano, T., Kato, T., Kosugi, T., Tsuneta, S., 
Watanabe, T., Kondo, I., Uchida, U. 1991, \solphys, 136, 10

\bibitem[Saito et al.(1994)]{Sai94}
Saito, T., Kozuka, Y., Tsuneta, S., and Minami, S.
1994, in {\it Proc. Int. Symp. on the Yohkoh Scientific Results, X-Ray
Solar Physics from Yohkoh}, ed. Y. Uchida, T. Watanabe, K. Shibata, \& 
H. S. Hudson (Tokyo: Universal Academy Press), 211

\bibitem[Saito et al.(2000)]{Sai00}
Saito, T., Shibata, K., Dere, K. P., Numazawa, S. 
2000, Adv. Space Res., 26, 807

\bibitem[Sheeley et al.(1975a)]{Shee75a}
Sheeley, N. R., Jr., Bohlin, J. D., Brueckner, G. E., Purcell, J. D.,
Scherrer, V., Tousey, R. 1975a, \solphys, 40, 103

\bibitem[Sheeley et al.(1975b)]{Shee75b}
Sheeley, N. R., Jr., Bohlin, J. D., Brueckner, G. E., Purcell, J. D.,
Scherrer, V. E., Tousey , R. 1975b, \apj, 196, L129

\bibitem[Shibata et al.(1994a)]{Shi94a}
Shibata, K., Nitta, N., Matsumoto, R., Tajima, T., Yokoyama, T., 
Hirayama, T., Hudson, H. 
1994a, in {\it Proc. Int. Symp. on the Yohkoh Scientific Results, X-Ray
Solar Physics from Yohkoh}, ed. Y. Uchida, T. Watanabe, K. Shibata, \& 
H. S. Hudson (Tokyo: Universal Academy Press), 29

\bibitem[Shibata et al.(1994b)]{Shi94b}
Shibata, K., Nitta, N., Strong, K. T., Matsumoto, R., Yokoyama, T., 
Hirayama, T., Hudson, H., Ogawara, Y. 
1994b, \apj, 431, L51

\bibitem[Takahashi et al.(1994)]{Taka94}
Takahashi, Ta., Saito, T., Shibata, K., Kozuka, Y., Minami, S., 
Mori, Y.
1994, in {\it Proc. Int. Symp. on the Yohkoh Scientific Results, X-Ray
Solar Physics from Yohkoh}, ed. Y. Uchida, T. Watanabe, K. Shibata, \& 
H. S. Hudson (Tokyo: Universal Academy Press), 305

\bibitem[Tousey et al.(1973)]{Tou73}
Tousey, R., et al. 1973, \solphys, 33, 265

\bibitem[Tsuneta et al.(1991)]{Tsu91}
Tsuneta, S., et al. 1991, \solphys, 136, 37

\bibitem[Verma(1998)]{Ver98}
Verma, V. K. 1998, J. Ind. Geophys. Union, 2, 65

\bibitem[Vourlidas et al.(1996)]{Vour96}
Vourlidas, A., Bastian, T. S., Nitta, N., Aschwanden, M. J. 
1996, \solphys, 163, 99

\bibitem[Wang(1998)]{Wang98}
Wang, Y.-M. 1998, \apj, 501, L145

\bibitem[Wang et al.(2006)]{Wang06}
Wang, Y.-M., Pick, M., \& Mason, G. M. 2006, \apj, 639, 495

\bibitem[Yokoyama \& Shibata(1995)]{Yoko95}
Yokoyama, T., \& Shibata, K. 1995, \nat, 375, 42

\bibitem[Yokoyama \& Shibata(1996)]{Yoko96}
Yokoyama, T., \& Shibata, K. 1996, \pasj, 48, 353
\end{thebibliography}
\end{document}